# Lattice Boltzmann prediction of transport properties in reconstructed nanostructures of organic matters in shales

Li Chen [a,b], Qinjun Kang [b], Lei Zhang [c], Wenquan Tao [a]

a: Key Laboratory of Thermo-Fluid Science and Engineering of MOE, School of Energy and Power Engineering, Xi'an Jiaotong University, Xi'an, Shaanxi 710049, China

b: Earth and Environmental Sciences Division, Los Alamos National Laboratory, Los Alamos, New Mexico, USA

c: School of Petroleum Engineering, China University of Petroleum, Qingdao, Shandong 266580, China

**Abstract**

Size, morphology and distributions of pores in organic matters of shale matrix are discussed based on high resolution images from experiments in the literature. 150 nanoscale structures of the organic matters are then reconstructed by randomly placing pore spheres with different diameters and overlap tolerances. Effects of porosity, the mean diameter and the overlap tolerance on void space connectivity and pore size distribution are studied. Further, a pore-scale model based on the Lattice Boltzmann method is developed to predict the Knudsen diffusivity and permeability of the reconstructed organic matters. The simulation results show that the mean pore diameter and overlap tolerance significantly affect the transport properties. The predicted Knudsen effective diffusivity is compared with Bruggeman equation and it is found that this equation underestimate the tortuosity. A modified Bruggeman equation is proposed based on the simulation results. The predicted intrinsic permeability is in acceptable agreement with Kozeny-Carman (KC) equation. In addition, a relationship is developed to determine the apparent permeability based on Knudsen diffusivity and intrinsic permeability. The predicted apparent permeability is compared with that predicted by various corrections in the literature. Knudsen's corrections match best with our numerical results and are recommended to calculate the apparent permeability.

# 1. Introduction

Gas-bearing shale reservoirs have become a major source of natural gas production in North American, and are expected to play an increasingly important role in Europe and Asia in the near future [1]. Advanced drilling techniques such as horizontal drilling and multi-stage hydraulic fracturing greatly promote the exploitation of hydrocarbon from shale reservoirs [2]. With hydraulic fracturing, both shale matrix and fracture systems exist inside the shale reservoirs. Consequently, four levels of void space can be identified in shale reservoirs including mesoscale hydraulic fractures, meso/microscale natural fractures, micro/nanoscale interparticle pores and nanoscale pores in organic matter [3].

Shale matrix is composed of predominantly clay minerals, quartz, pyrite and organic matter [4-8]. The organic matter is the source of shale gas (methane) and plays an important role in shales. It adsorbs shale gas as well as stores free shale gas (shale gas is also stored in other void space mentioned above). The weight percentage of the organic matter in shales, called total organic content (TOC), is an important indicator of the gas reserve. Pores are formed when oil and gas were generated, and are likely to be related to the mature of the shales. There pores are dominantly of nanosize, with pore diameters in the range of a few nanometers to hundreds of nanometers [5]. In some shale plays, most pores in the shale matrix are associated with organic matters, and the permeability of the organic matter is of great importance for shale gas transport from the matrix to the fracture network. As a separate porous medium in shale matrix which is distinct from surrounding nonorganic matters, organic matter greatly affects the petrophysical properties of shale matrix such as density, porosity and wettability [7].

While shale gas flow in the fractures is very important for the production, how the shale gas transports inside the shale matrix and how it migrates from shale matrix to the fractures also play important roles. A fundamental understanding of multiple physiochemical processes in shale matrix is crucial for improving the gas production and lowering production costs. Multiple transport mechanisms play roles during shale gas extraction due to the widely spanning length scale [9, 10]. Because of the nanoscale characteristics of the shale matrix, Darcy law, which is widely adopted in the conventional formation, cannot realistically describe the variety of the relevant flow regimes other than the viscous flow regime [3, 9, 11, 12]. The dominant mechanisms of shale gas transport in shale matrix is slip flow and Knudsen diffusion [9].

To date, petrophysical properties of shale matrix remain poorly understood, which is of significant importance for revealing the transport mechanisms as well as economically assessing the storage of shale gas in shale reservoirs [7]. Advanced experimental techniques have been employed/developed to observe the nanoscale structures of shale matrix [5-8]. Particularly, scanning electron microscopy (SEM) combined with focused ion beam (FIB) milling allows one to directly visualize the nanoscale structures of shales on high quality flat surfaces [7]. It has been revealed that TOC, pore morphology, porosity and pore size can differ remarkably between shale plays and even in the same play [4-8]. This scenario further complicates the investigation of petrophysical properties of shale matrix.

The objective of our study is to thoroughly investigate the physicochemical processes during the extraction of shale gas using a multi-scale modeling strategy. For the first step study, the work in the present study aims to investigate the geometrical properties (porosity, pore size distribution and pore connectivity) and macroscopic effective transport properties (permeability, effective diffusivity) of the organic matter at the nanoscale. A nanoscale model based on the lattice Boltzmann method (LBM) is developed to investigate the fluid flow and mass transport processes in reconstructed structures of the organic matters. Effective transport properties including permeability and effective diffusivity are predicted, which will be upscaled into the next step study where microscale transport processes in shale matrix including organic and nonorganic matters are investigated. The results from the microscale study will be further upscaled into a mesoscale study in which transport processes in the fracture systems are explored.

The present work develops a nanoscale numerical model based on the LBM to investigate the transport processes in organic matters. The remaining of the present work is arranged as follows. We first describe the reconstruction processes of the geometries of the organic matters based on the experimental observations in the literature in Section 2. The numerical method of predicting effective transport properties is introduced in Section 3. Then, in Section 4 the structure characterizations of the reconstructed structures are performed including porosity, pore size distribution and void space connectivity. Predicted effective Knudsen diffusivity and intrinsic permeability are also presented in this section. Then apparent permeability is determined based on the predicted transport properties, and are compared with existing corrections in the literature. Finally some conclusions are drawn in Section 5.

**2 Reconstruction of the organic matter**

The reconstruction of 3D porous media can be obtained by two approaches: the computer tomography using experimental techniques such as SEM combined with FIB, and virtual reconstruction algorithm. The former scans the porous media to obtain images of several sections and then integrates these images to achieve the geometry construction [7]. The later reconstructs the porous structures based on the statistical information of the porous medium. Due to low cost and easy implementation of geometry generation, the virtual reconstruction algorithm has been widely employed to reconstruct porous medium; it also offers a convenient way to perform systematic investigations of the effects of geometrical properties of the porous medium, such as porosity, pore size distribution, connectivity on petrophysical properties.

Recently, a great effort has been devoted to experimentally understand the geometrical characteristics of shale matrix [4-8, 13-15]. Compared with indirect measurement techniques such as Mercury injection porosimetry (MIP), Gas adsorption BET and Nuclear magnetic resonance (NMR), direct measurement techniques can provide visual images of the pore morphologies and distributions. Particularly, very recently SEM combined with FIB milling allows visualize the nanoscale structures of shales on high quality flat surfaces [7]. The advanced techniques present new insights in the nanoscale structures of the kerogens including concentration, morphology, porosity, pore size and pore shapes. Based on the experimental observations [4-8, 13-15], the following conclusions of the structures of the organic matter can be obtained. Organic matter usually presents as discrete grains in the shale matrix surrounded by nonorganic matters. Nanopores are very common in the organic matters, which are formed during the generation of oil and gas. The void space structures of organic matters are really complex, which are affected by several factors including maturity, organic composition and late localized compaction [4, 5]. Even answers to porosity are elusive. Loucks et al. reported a range of porosity 0%~30% in the organic matter of Barnett shales, North Texas [5], while Sondergeld et al. [7] reported a porosity of 50%. The morphology of the nanopores in the organic matter is diverse, including round, ellipsoidal, triangular, prolate, polyhedral and irregular shapes [5, 7]. Nanopores are generally nonequant to some degree [5]. The pore size is in a wide range, from a few nanometers to hundreds of nanometers [5-8]. The distributions of the pores also cannot be generalized, nevertheless in most cases the pores are quite randomly and uniformly distributed [5, 7].

In the present study, we assume that the void space in the kerogen can be described by uniformly distributed spheres. This assumption is based on the experimental observations (Fig. 19 in [7]; Fig. 5 and Fig. 6 in [5]; Fig. 9 in [8]). The general scheme to reconstruct the organic matter is to randomly place pores with spherical shape into the computational domain. Certain parameters required to be specified before the reconstructions are porosity $\varepsilon$, the radius of each sphere $d$, the probability that the spheres are required to overlap $p_0$ and the overlap tolerance $\xi$. The overlap tolerance, defined as the volume of the overlapping region of a sphere with another sphere to its total volume, determines the maximum overlapping volume. The structure generation algorithm is described as follows [16]:

1) The first pore sphere is generated in the computational domain. Random number generators are employed to generate its centers and diameter;

2) A random number generator is employed to determine whether or not the next sphere is allowed to overlap with existing pore spheres in the computational domain. Further, a random number generator is used to generate sphere centers and diameter of a trial pore. a) If the pore sphere is required to overlap with existing spheres, it must overlap with an existing sphere and the overlapping portion should be less or equal to the overlap tolerance; b) otherwise, it can be separated from other spheres, or if it overlaps with any existing spheres, the overlapping portion cannot exceed the overlap tolerance;

3) Repeating Step 2 until the specified porosity has been achieved.

The diameter of pore spheres follows a uniform distribution ($<d>$-15nm, $<d>$+15nm), with $<d>$ the mean diameter chosen as 30nm, 45 nm and 60 nm. The porosity ranges between 0.1 and 0.55, with an interval of 0.05, based on the experimental results of Loucks et al. [5] and Sondergeld et al. [7]. $\xi$ is set as 0.1, 0.2 and 0.3, respectively. The structure reconstruction algorithm relies on a random process. Therefore, the geometry as well as the properties might vary from one generated structure to another even if the input parameters are identical. To consider such variability, therefore, it is necessary to generate several geometries from the same input parameters. In the present study, five structures are generated for each configuration. Consequently, totally 150 nanoscale structures of organic matters are reconstructed. A 3D representation of a geometry is given in Fig. 1 as an example with $<d>$=45nm, $\xi$=0.2 and $\varepsilon$=0.4. The computational domain is a cube with size of $L_x \times L_y \times L_z$ with $L_x=L_y=L_z$=300nm, and is discretized by 200×200×200 lattices, leading to a resolution of 1.5nm per lattice. As shown in Fig.1, the pore spheres with different sizes are randomly distributed in the domains, and most of them are overlapped with each other, Loucks et al. [5] experimentally observed that narrow throats with diameter peak of 10~15nm between larger pores help to connect the void space (Fig. 6 in [5]). High resolution Backscattered electrons (BSE) images of a region of kerogen in Horn River provided by Curtis et al. [8] (Fig. 9 in [8]) also shows that the organic matter contains both large and small pores; and large pores are perforated with numerous smaller pores. The overlapping regions with small pore size in our reconstruction scheme can be considered as "narrow throats" or "small pores" which connect the large pore spheres. Obviously, such overlapping is critical for the connection between pore spheres as well as the transport processes in the organic matter.

## 3. Numerical method

Pore-scale models directly describe the transport processes in porous media based on the realistic structures. They can be used to improve the fundamental understanding of the transport processes in the nanopores of organic matters. Owing to its excellent numerical stability and constitutive versatility, the LBM has developed into a powerful technique for simulating transport processes and is particularly successful in transport processes applications involving interfacial dynamics and complex geometries, such as multiphase flow and flow in porous media [17-19]. The LBM considers flow as a collective behavior of pseudo-particles residing on a mesoscopic level, and solves Boltzmann equation using a small number of velocities adapted to a regular grid in space. The LBM has been successfully applied to a wide range of complex transport problems [18], such as porous flow [20], multiphase flow [17, 20], particle flow [21], reactive transport processes [22-27], etc. In the present study, intrinsic permeability and effective Knudsen diffusivity of the organic matters are numerically predicted using the LBM.

3.1 LB fluid flow model

The most popular LB fluid flow model is the Bhatnagar-Gross-Krook (BGK) model (BGK) or single relaxation time (SRT) model [28]. However, a viscosity-dependent permeability is usually obtained when adopting SRT LBM for simulating fluid flow [29]. In order to overcome such defect, the multi-relaxiton-time (MRT) model has been proposed recently to simulate fluid flow through porous media [29, 30]. The MRT-LBM model transforms the distribution functions in the velocity space of the SRT-LBM model to the moment space by adopting a transformation matrix. In the SRT-LBM model, the evolution equation for the distribution functions is as follows

$$f_i(\mathbf{x}+\mathbf{e}_i\Delta t,t+\Delta t) - f_i(\mathbf{x},t) = \mathbf{S}[f_i^{eq}(\mathbf{x},t) - f_i(\mathbf{x},t)] \quad i = 0 \sim N \tag{1}$$

where $f_i(\mathbf{x},t)$ is the $i$th density distribution function at the lattice site $\mathbf{x}$ and time $t$. $\mathbf{S}$ is the relaxation matrix. For the D3Q19 (three-dimensional nineteen-velocity) lattice model with $N=18$ used in this work as shown in Fig. 2, the discrete lattice velocity $\mathbf{e}_i$ is given by

$$\mathbf{c}_i = \begin{cases} 0 & i = 0 \\ (\pm 1,0,0), (0,\pm 1,0), (0,\pm 1,0) & i = 1 \sim 6 \\ (\pm 1,\pm 1,0), (0,\pm 1,\pm 1), (\pm 1,0,\pm 1) & i = 7-18 \end{cases} \tag{2}$$

$f^{eq}$ is the $i$th equilibrium distribution function and is a function of local density and velocity

$$f_i^{eq} = w_i \rho \left[ 1 + \frac{\mathbf{e}_i \cdot \mathbf{u}}{(c_s)^2} + \frac{(\mathbf{e}_i \cdot \mathbf{u})^2}{2(c_s)^4} - \frac{\mathbf{u} \cdot \mathbf{u}}{2(c_s)^2} \right] \tag{3}$$

with the weight coefficient $w_i$ as $w_i=1/3$, $i=0$; $w_i=1/18$, $i=1,2,\ldots,6$; $w_i=1/36$, $i=7,8,\ldots,18$. $c_s = 1/\sqrt{3}$ is the speed of sound. By multiplying a transformation matrix $\mathbf{Q}$ ( a $(N+1)\times(N+1)$ matrix) in Eq. (1), the evolution equation in the moment space can be expressed as

$$\mathbf{m}(\mathbf{x}+\mathbf{c}\Delta t,t+\Delta t) - \mathbf{m}(\mathbf{x},t) = \hat{\mathbf{S}}[\mathbf{m}^{eq}(\mathbf{x},t) - \mathbf{m}(\mathbf{x},t)] \tag{4}$$

where

$$\mathbf{m} = \mathbf{Q} \cdot \mathbf{f}, \quad \mathbf{m}^{eq} = \mathbf{Q} \cdot \mathbf{f}^{eq}, \quad \hat{\mathbf{S}} = \mathbf{Q} \cdot \mathbf{S} \cdot \mathbf{Q}^{-1} \tag{5}$$

with $\mathbf{m}$ and $\mathbf{m}^{eq}$ as the velocity moments and equilibrium velocity moments, respectively. $\mathbf{Q}^{-1}$ is the inverse matrix of $\mathbf{Q}$, both of which are given in the Appendix A. The transformation matrix

**Q** is constructed based on the principle that the relaxation matrix $\hat{\mathbf{S}}$ (a $(N+1)\times(N+1)$ matrix) in moment space can be reduced to the diagonal matrix [30], namely

$$\hat{\mathbf{S}} = \mathrm{diag}(s_0, s_1, \ldots, s_{17}, s_{18}) \tag{6}$$

with [29]

$$s_0 = s_3 = s_5 = s_7 = 0,\ s_1 = s_2 = s_{9-15} = \frac{1}{\tau},\ s_4 = s_6 = s_8 = s_{16-18} = 8\frac{2\tau-1}{8\tau-1} \tag{7}$$

$\tau$ is related to the fluid viscosity by

$$\tau = \frac{\upsilon}{c_s^2 \Delta t} + 0.5 \tag{8}$$

The equilibrium velocity moments $\mathbf{m}^{eq}$ are given in Appendix A. The density and momentum are determined by

$$\rho = \sum_i f_i,\ \mathbf{j} = \sum_i f_i \mathbf{e}_i \tag{9}$$

Eq. (4) and Eq. (A) can be recovered to Navier-Stokes equations using Chapman–Enskog multiscale expansion under the low Mach number limitation. In the present study, the main steps for applying the MRT-LBM are as follows: 1) calculating the equilibrium velocity moments using Eq. (A); 2) evolution the moments according to Eq. (4); 3) transforming the moments back to the distribution function in velocity space using Eq. (5); 4) Implementing the stream step and boundary conditions using the distribution function in velocity space; 5) calculating the density and velocity using Eq. (9). Therefore, the collision step is implemented in the moment space while the stream step is carried out in the velocity space.

3.2 LB mass transport model

For pure methane diffusion in the organic matter, the evolution equation for the concentration distribution function is as follows

$$g_i(\mathbf{x}+\mathbf{e}_i \Delta t, t+\Delta t) - g_i(\mathbf{x},t) = -\frac{1}{\tau_g}(g_i(\mathbf{x},t) - g_i^{eq}(\mathbf{x},t)) \tag{10}$$

where $g_i$ is the distribution function. In coincidence with the 19 direction labeling algorithm introduced in Section 3.4, D3Q19 lattice model is adopted. It is worth mentioning that for simple geometries, D3Q7 lattice model (or D2Q5 model in 2D) is sufficient to accurately predict the diffusion process and properties, which can greatly reduce the computational resources, compared with D3Q19 (or D2Q9 in 2D), as proven by our previous work [20, 22, 31-34].

However, for complex porous structures, especially for those with low porosity, such as shales, using reduced lattice model will damage the connectivity of the void space, thus leading to underestimated effective diffusivity. The equilibrium distribution function $g^{eq}$ is defined as $g_i^{eq} = C_{CH_4}/a_i$ with $a_0$=1/3, $a_{1\sim6}$=1/18 and $a_{7\sim18}$=1/36. The concentration and the Knudsen diffusivity $D_K$ are obtained by $C_{CH_4} = \sum g_i$, $D_K = \frac{1}{3}(\tau_g - 0.5)\Delta x^2 / \Delta t$. When the character length of a system is comparable to or smaller than the mean free path of the molecules involved, collisions between a molecule and the solid wall are more frequent than that between molecules, and such mean of diffusion is called Knudsen diffusion. The Knudsen diffusivity in a local pore with pore diameter $d_p$ (m) is calculated by the following equation according to the dusty gas model [35]

$$D_K = \frac{d_p}{3}\sqrt{\frac{8RT}{\pi M}} \tag{11}$$

where $R$ is the gas constant, $T$ the temperature and $M$ the molar mass. According to Eq. (11), the Knudsen diffusivity is a function of local pore diameter. A reference pore size $d_{p,ref}$=60nm is chosen, and the corresponding relaxation time $\tau_{g,ref}$ is set as 1.0. Hence, for any local pore with size $d_p$, the relaxtion time can be calculated as

$$\tau_g = \frac{d_p}{d_{p,ref}}(\tau_{g,ref} - 0.5) + 0.5 \tag{12}$$

Through Eq. (12), the effects of pore size distributions on the Knudsen diffusion process inside the nanoscale structures of porous shales are taken into account. In the present study, $d_p$ is an effective pore diameter calculated by the 13 direction averaging scheme introduced in Section 4.1. Eq. (10) combined with the equilibrium distribution function can be proved to recover the pure diffusion equation using Chapman–Enskog multiscale expansion [31, 32].

3.3 Boundary condition

Unlike conventional numerical methods where the fundamental variables are macro physical quantities such as density, velocity, temperature and concentration, in the LB framework, the fundamental variables are the distribution functions. Therefore, boundary conditions based on the distribution functions are required when performing LB simulations.

As introduced in Section 3.1, the collision step is implemented in the moment space while the stream step is carried out in the velocity space. Therefore, the boundary conditions developed in SRT model can still be applied to MRT-LBM. In the present study, for fluid flow three kinds of boundary conditions are adopted: no-slip boundary condition at the fluid-solid interface inside the domain, pressure boundary condition at the inlet and outlet of the computational domain (*x* direction), and periodic boundary conditions for the other four boundaries (*y* and *z* directions). In the LB framework, for no slip boundary condition, the half way bounce-back scheme is employed; for periodic boundary conditions, the unknown distributions at one boundary (*y*=0 for example) is set to that at the other boundary (*y*= $L_y$, for example); and for pressure boundary condition, the non-equilibrium extrapolation method proposed by Guo et al. [36] is adopted for its good accuracy. It should be noted that the pressure difference between inlet and outlet should be small enough, one reason to ensure low Mach number fluid flow for the LB model and the other reason to eliminate the inertial effect.

For methane diffusion, four types of boundary conditions are used: the no-flux boundary condition at the fluid-solid interface inside the domain, concentration boundary condition at the inlet and outlet of the computational domain (*x* direction), and periodic boundary conditions for the other four boundaries (*y* and *z* directions). In the LB framework, for the concentration boundary condition, the unknown distribution function is determined using the equilibrium distribution functions; for the non flux boundary condition, bounce-back scheme is adopted.

3.4. Intrinsic permeability and effective Knudsen diffusivity

Intrinsic permeability of a porous medium only depends on the porous structures and is not affected by the fluid type and flow condition. It can be obtained by simulating the fluid flow in the organic matter without considering the slip on the solid surface. When the simulation of fluid flow reaches steady state using the LB fluid flow model, the intrinsic permeability is calculated based on the Darcy's Law [37]

$$J_\mathrm{d} = -\frac{\rho k_\mathrm{d}}{\mu}\nabla p \tag{13}$$

where the flow flux $J_\mathrm{d}$ equals to $\rho <u>$ with $<u>$ as the volume-averaged fluid velocity along flow direction.

The effective Knudsen diffusivity $D_{K,\mathrm{eff}}$ along the *x* direction is calculated by

$$D_{K,\mathrm{eff}} = \frac{(\int_0^{L_y}\int_0^{L_z}(D_K\frac{\partial C}{\partial x})|_{L_x} dydz)/L_y L_z}{(C_\mathrm{in}-C_\mathrm{out})/L_x} \tag{14}$$

where $C_\mathrm{in}$ and $C_\mathrm{out}$ is the inlet and outlet concentration, respectively. $|_{L_x}$ means local value at *x*=$L_x$.

3.5 Validation

For validation of the LB fluid flow model, simulation is performed for flow through an 3D open cube in which equal-sized spheres of diameter $d$ are arranged in periodic BCC (body-centered cubic) arrays and $Re$ number is much less than unity, as shown in Fig. 3. 100×100×100 lattice is employed, with pressure difference applied on the left and right boundaries and periodic boundary conditions on the other four directions. Fig. 3(a) shows the simulated streamlines inside the periodic BCC arrays with a porosity of 0.915. A comparison between the simulated permeability and Kozeny-Carman (KC) equation [38] for the BCC arrays with different porosity is shown in Fig. 3(b). The KC equation, a widely used semi-empirical equation for predicting permeability, is given by $k_d = A\varepsilon^3/(1-\varepsilon)^2$, with $A$ is the KC constant and is set as $A = (d^2/180)$ for packed-spheres porous media. Analytic solutions from [39] are also displayed in Figure. 3(b). As shown in the figure, the MRT-LBM simulation results agree well with the analytical solution for the whole range of $\varepsilon$. However, the KC equation presents upward discrepancy and downward discrepancy for high and low porosity, respectively.

Further, the LB mass transport model is validated by predicting the effective diffusivity in a cubic domain containing a sphere whose diameter is the same as the side length of the cube [16]. For such configuration, the porosity is always 1-$\pi$/6. 100×100×100 lattices is employed and the predicted effective diffusivity is 0.3317, in good agreement with the results in [16].

## 4. Results and discussion

In the LBM framework, the simulation variables are in the lattice units instead of physical units. The physical variables such as permeability $k_p$ and diffusivity $D_p$ (subscript "p" denotes "physical space") can be calculated from the quantities in lattice system (subscripted by L) by the following expression

$$\frac{D_P}{\Delta x_P^2/\Delta t_P} = \frac{D_L}{\Delta x_L^2/\Delta t_L}, \quad \frac{k_P}{l_P^2} = \frac{k_L}{l_L^2} \tag{15}$$

$\Delta x_P$ and $\Delta t_P$ equal unit in the LBM. After the computational domain is defined and discretized, the physical length of a lattice $\Delta x_P$ can be determined.

4.1 Characteristics of the nanoscale structures

Porosity and pore size distribution (PSD) of the organic matters are critical for both gas storage and flow quantification. Porosity studied ranges in 0.1~0.55, as mentioned in Section 2 based on the experiments [5, 7]. A 13 direction averaging method [16] is adopted to determine the PSD of the reconstructed organic matter. In this method, 13 directions are 3 directions along the x, y and z axis, 6 diagonal directions in x-y, x-z and y-z planes with 2 diagonal directions in each plane, and 4 diagonal directions traversing the 3D space. For each pore cell in a certain direction, a pore length, which is defined as the length of a line occupied by pore cells only, is determined. The pore size of the current pore cell is calculated as the averaged value of the pore lengths in the 13 directions. Fig. 4 shows the PSD for the reconstructed structures with $<d>$=45nm and $\xi$=0.1 for different porosities. The PSD of all the cases is unimodal and can be approximately described as a normal distribution. Both the mean and the standard deviation go up as the porosity increases. For all the reconstructed structures with different porosities, the pore size is located in the range of a few nanometers to hundreds of nanometers, indicating the nanoscale characteristics of reconstructed organic matters.

Shale gas transports through the connected pores in the organic matter. Connectivity of the void space is therefore very important. For the reconstructed structures of the organic matters, "transport" and "dead" pore cells are distinguished. Here, a pore cell is "transport" means that this cell belongs to a continuous percolation path throughout the entire organic matter. If a portion of void phase does not penetrate the entire organic matter, shale gas entering the shale from one end cannot arrive at the other end through this portion, and this portion is called "dead". For identifying the "transport" and "dead" portions in the reconstructed organic matter, a connected phase labeling algorithm is developed. This algorithm scans the entire domain, checks the connectivity of a cell with its neighboring 18 points, labels the cell depending on the local connectivity, and finally assigns any portion with a distinct label if it is disconnected from other portions in the domain. Here, neighboring 18 points are used, in coincidence with the D3Q19 lattice model in the LBM framework adopted in the present study. The connectivity of void space is defined as the ratio of the number of "transport" cells to the total pore cells. Fig. 5(a) plots the connectivity of the reconstructed structures for different porosities with $\xi$=0.1. First, the connectivity increases as the porosity increases. Under lower porosity, there are fewer spheres in the organic matters which are sparsely distributed and are less likely to be connected with each other. Second, the discrepancy of the connectivity between different samples of the same porosity becomes larger as the porosity reduces. For example, for the cases with $<d>$ =45nm, when the porosity is lower than 0.15, all the pores are dead for the reconstructed structures, while for the porosity greater than 0.4, the connectivity is almost unit, meaning all the pores are "transport". This is because fewer pores under lower porosity result in larger fluctuation of the void space. Finally, it can be seen that higher diameter results in lower connectivity and larger discrepancy of the connectivity. The pores become to be connected when the porosity is greater than 0.15 for $<d>$ =30nm, while for the structures with $<d>$ =60nm, the minimum porosity rises to 0.25. This is because under the same porosity, there are less pore spheres in the organic matters for a higher diameter. Fig. 5(b) displays the effects of overlapping tolerance $\xi$, where all the cases are with $<d>$=45nm. It can be seen that as $\xi$ increases, the connectivity decreases and the discrepancy of the connectivity goes up. This is because the pore spheres tend to aggregate under higher overlap tolerance. On the whole, the results indicate that to obtain better inner-connected void space in the organic matters, both the pore size and the overlapping tolerance should be lower, especially under low porosity.

4.2 Effective Knudsen diffusivity

Fig. 6 shows the concentration distribution within the organic matter with mean diameter of 45nm, porosity of 0.4 and overlap tolerance of 0.2. Based on the concentration filed obtained, the effective diffusivity is calculated using Eq. (14). Fig. 7(a) shows the effective diffusivity vs porosity for organic matters with pore spheres of different mean diameter $<d>$. In the figure, the overlap tolerance is 0.1 and the diffusivity is normalized by reference diffusivity, $D_{K,ref} = 1.31 \times 10^{-5}$ m s$^{-2}$, which is the Knudsen diffusivity with $d$=60nm and $T$=323 according to Eq. (11). Clearly, as the porosity increases, the diffusivity increases. The mean diameter greatly affects the effective diffusivity. Generally, a larger mean diameter generates higher diffusivity. This is because larger spheres results in higher local Knudsen diffusivity. For the same porosity, larger pore spheres lead to higher fluctuation of the structures, thus result in higher discrepancy of diffusivity. Fig. 7(b) displays the effects of overlap tolerance on the effective Knudsen diffusivity, where the mean diameter $<d>$=45nm. The overlapping region is crucial for the connection of void space as well as the transport processes. As can be seen from Fig. 7(b), a higher overlap tolerance can enhance the mass transport. This is because the local transport within the overlapping region is the limiting factor for that in the organic matter, because the local pore size is smaller compared with that in the bulk pore spheres. The lower the overlap tolerance is, the smaller the overlapping region is, and hence the lower the local diffusivity is. The discrepancy for samples with the same input parameters increases as the overlap tolerance rises, which is because the pore spheres are more likely to aggregate with larger overlapping tolerance. Finally, the predicted Knudsen diffusivity ranges from 10$^{-9}$ m$^2$ s$^{-1}$ to 10$^{-6}$ m$^2$ s$^{-1}$. To the best of our knowledge, there are no experiments particularly devoted to measure the diffusivity within the organic matters. Javadpour et al. reported a diffusivity of $4 \times 10^{-7}$ m$^2$ s$^{-1}$ in a shale sample [3] (The TOC and the distributions of the organic matters in the shale sample is not mentioned). Their measured value is in the range of our numerically predicted value.

Macroscopic models of transport processes in shale matrix highly depend on empirical relationships between macroscopic transport properties and statistical structural information of porous components (permeability vs porosity, diffusivity vs porosity). Bruggeman equation has been widely used in the macroscopic models to calculate the effective diffusivity

$$D_{eff} = D \frac{\varepsilon}{\tau} \qquad (16)$$

where $\tau$ is the tortuosity of a porous medium, and is set as $\tau=\varepsilon^{-0.5}$ in Bruggeman equation, leading to effective transport properties as a pure function of the volume fraction, namely, $D_{eff}=D\varepsilon^{\alpha}$ with α=1.5 [40]. Although Bruggeman equation has been widely adopted, it was determined empirically from sphere packed porous medium with high porosity, which therefore cannot reflect the complex structures of shales with lower porosity. As can be seen in Fig. 7, the Bruggeman equation greatly overestimates the effective diffusivity. Following the relationship $D_{eff}=D\varepsilon^{\alpha}$, it is found that α ranges 2.8~4.0 for the organic matters with different pore diameter and overlap tolerance based on our nanoscale simulation results; and the lower the porosity and the overlap tolerance are, the higher the α. This implies as the porosity and overlap tolerance decrease, the void space within the organic matter becomes more tortuous. Bruggeman equation greatly underestimates the tortuosity in organic matter, especially for lower porosity. Our numerical simulation results show that in the future macroscopic modeling, a higher α in the range of 2.8~4.0 is recommended to use in Bruggeman equation. Fig. 7 also shows the revised Bruggeman equation with α=2.8 and 4.0, respectively. As shown in the Figure, they are in good agreement with values predicted by our simulations.

4.3 Intrinsic Permeability

Fig. 8 shows pressure and velocity vector distributions within the reconstructed organic matter with mean diameter of 45nm, porosity of 0.4 and overlap tolerance of 0.2. Fluid pathway within the porous structures is very tortuous, and in the overlap region flow is relatively strong. Fig. 9(a) shows the intrinsic permeability vs porosity for organic matters with pore spheres of different mean diameter <d>, where the overlap tolerance is 0.1, while Fig. 9(b) shows that for different tolerance with <d>=45 nm. Similar to that of effective diffusivity, the intrinsic permeability increases as the porosity, mean pore diameter and overlap tolerance are increased. The intrinsic permeability predicted ranges $10^{-21}(10^{-6}mD) \sim 10^{-17}(10^{-2}mD)$ m². Also, there are no experiments particularly devoted to measure the permeability within the organic matters. Javadpour et al. [3] measured 152 shale samples from nine reservoirs, and reported that a permeability peak at around $5.4\times10^{-5}$mD, and the permeability of 90% samples is less than $1.5\times10^{-4}$mD. Wang and Reed [6] plotted together the permeability of shales in north American with porosity of 0~0.1, and found that the permeability ranges from $1\times10^{-12}$mD to $1\times10^{-2}$ mD. Our predicted intrinsic permeability of the organic matter is consistent with that from the literature.

The most adopted empirical relationship for predicting permeability is the KC equation [38]

$$k_d = A\frac{\varepsilon^3}{(1-\varepsilon)^2} \qquad (17)$$

with $A$ is the KC constant and is set as $d^2/180$ for packed-spheres porous media, where $d$ is the diameter of the solid spheres. Note that in the present study, the spheres represent the void space rather than the solid phase; nevertheless, in both cases, the permeability will increase as $d$ increase for the same porosity. Hence, in the present study, Eq. (17) is still employed to predict the intrinsic permeability of the reconstructed organic matters with $d$ chosen as the mean diameter <$d$> of the pore spheres. As shown in Fig. 9, the value of permeability predicted by KC equation presents acceptable agreement with that obtained from our nanoscale simulations, especially for lower porosity. In Fig. 9(a), as porosity increases, the discrepancy between them becomes larger. In Fig. 9(b), the KC equation is in best agreement with the case with overlap tolerance of 0.3.

4.4 Apparent permeability

4.4.1 Overview of existing corrections

Gas slippage phenomenon occurs when the mean free path of the gas is comparable to the character length of the domain. Klinkenberg [41] first conducted the study of gas slippage in porous media. It was found that the permeability of gas through a tight porous media is higher than that of liquid, which is due to the slip of gas molecules on the solid surface. Klinkenberg [41] proposed a linear correlation for correcting the gas permeability

$$k_a = k_d f_c, f_c = (1+\frac{b_k}{P}) \tag{18}$$

where $f_c$ is the correction factor and $P$ is the pressure. $k_d$ is called Klinkenberg's corrected permeability, which is the permeability of liquid, or the intrinsic permeability predicted in Section 4.3. $k_d$ only depends on the porous structures of a porous medium and is not affected by the operating condition and fluid properties. $b_k$ is the Klinkenberg's slippage factor, which depends on the molecular mean free path $\lambda$, character pore size of the porous media $r$ and the pressure [41]

$$\frac{b_k}{P} = \frac{4c\lambda}{r} \approx 4Kn, c \approx 1 \tag{19}$$

where $Kn$ is Knudsen number, defined as the ratio between gas mean free path to the pore size. Klinkenberg's correction is a first-order correction. Beskok and Karniadakis [42] developed a second-order correction, and they showed that this correction is valid for the entire $Kn$'s range, which is expressed as

$$f_c = (1+\alpha(Kn)Kn)\left[1+\frac{4Kn}{1-bKn}\right] \tag{20}$$

with $b$ is the slip coefficient and equal to -1 for slip flow. $\alpha(Kn)$ is the rarefaction coefficient. The expression of $\alpha(Kn)$ in [42] is very complex. Civan [43] proposed a much simplified expression for $\alpha(Kn)$

$$\alpha(Kn) = \frac{1.358}{1 + 0.170 Kn^{-0.4348}} \tag{21}$$

Eq. (20) combined with Eq. (21) is called Beskok and Karniadakis-Civan's correction in the present study.

Based on Eq. (18), various expressions of $b_k$ have been proposed in the literature, as listed in Table 1. Heid et al. [44] and Jones and Owens [45] proposed similar expressions relating $b_k$ to $k_d$. Sampath and Keighin [46] and Florence et al. [47] developed a different form by relating $b_k$ to $k_d/\varepsilon$.

Table 1 Various corrections in the literature

| Literature | $b_k$ | units |
|---|---|---|
| Heid et al., 1950, [44] | $11.419(k_d)^{-0.39}$ | |
| Jones and Owens, 1980 [45] | $12.639(k_d)^{-0.33}$ | |
| Sampath and Keighin, 1982 [46] | $13.85(k_d/\varepsilon)^{-0.53}$ | $b_k$: psi, $k_d$: mD |
| Florence et al., 2007 [47] | $\beta(k_d/\varepsilon)^{-0.5}$<br>Membrane: $\beta=32.766$ | |

4.4.2 The dusty gas model

As the pore size of porous medium decreases, or $Kn$ increases, the application of Darcy law is questionable, because the boundary slip phenomenon and the Knudsen diffusion begin to play important roles [3]. In shale matrix, the transport behaviors of shale gas are combined results of viscous flow and Knudsen diffusion [3, 11, 12]. Apparent permeability $k_a$, which can account for both transport mechanisms, is defined and used in Darcy law (Eq. (13))

$$J = -\frac{\rho k_a}{\mu} \nabla p \tag{22}$$

with $J$ equal

$$J = J_d + J_k \tag{23}$$

where $J_d$ is given by Eq. (13). Subscript "k" represents "Knudsen". $J_k$ is the Knudsen diffusion flux term and is expressed as

$$J_k = -MD_{k,eff}\nabla C = -MD_{k,eff}\nabla(\frac{p}{zRT}) = -\frac{\rho}{p}D_{k,eff}\nabla p \tag{24}$$

$z$ is the gas compressibility factor, accounting for the effect of non-ideal gas. $D_{k,eff}$ (m$^2$ s$^{-1}$) is the effective Knudsen diffusivity and $D_k$ is given by Eq. (11). Therefore, the total flow flux can be expressed as

$$J = J_d + J_k = -\frac{\rho k_d}{\mu}\nabla p - \frac{\rho}{p}D_{k,eff}\nabla p = -k_d(1+\frac{D_{k,eff}\mu}{pk_d})\frac{\rho}{\mu}\nabla p \tag{25}$$

Eq. (25) is same to the single component gas transport model in a porous medium derived based on the dusty gas model (DGM) [48, 49]. Shale gas is a mixture of several components, but the methane dominates with mole fraction of about 90% [3]. Therefore, shale gas can be considered of pure methane, and thus binary gas diffusion is neglected. Based on Eq. (25), the apparent permeability can be determined

$$k_a = k_d f_c, \quad f_c = (1+\frac{D_{k,eff}\mu}{k_d P}) \tag{26}$$

4.4.3 Comparisons between different corrections

Here, the correction factor predicted by Eq. (26) is compared with that by Klinkenberg's correction (Eq. (18)) and Beskok and Karniadakis-Civan's correction (Eq. (20)). A simple case of fluid flow and Knudsen diffusion in a cylinder with diameter $d_p$ is considered. For such transport problem, the intrinsic permeability is

$$k_d = \frac{(d_p/2)^2}{8} \tag{27}$$

The effective diffusivity is given by Eq. (11). Substituting Eqs. (11) and (27) into Eq. (26) leads to

$$f_c = (1+\frac{64}{3\pi}Kn) \tag{28}$$

where the following definition of mean free path $\lambda$ is adopted [50]

$$\lambda = \frac{\mu}{p}\sqrt{\frac{\pi RT}{2M}} \tag{29}$$

Fig. 10 shows the correction factor in the cylinder predicted by Klinkenberg's correction and Beskok and Karniadakis-Civan's correction. When $Kn$ is quite low, i.e., in the range of Darcy flow regime ($Kn<0.01$), the above equations predict quite similar values as the term with $Kn$ in Eq. (18), (20) and (28) can be neglected. However, as $Kn$ increases, the discrepancy between Klinkenberg's correction and other corrections becomes large. The Beskok and Karniadakis-Civan's correction is more accurate than Klinkenberg's correction, especially at high $Kn$ regime [51]. Values predicted by Eq. (28) agree well with Beskok and Karniadakis-Civan's correction. This agreement leads to the theoretical basis of the present study.

In the literature, to determine the apparent permeability of microchannels or micro porous media, in which the fluid flow regime falls in slip flow or transition flow regime, Navier-stokes equation is solved with modified local viscosity and modified slip flow boundary conditions [52, 53]. Delicate numerical schemes are required in such simulations. Through such simulations, the detailed distribution of fluid flow, pressure as well as the apparent permeability can be obtained. However, for reservoir simulations, the empirical relationships (permeability-porosity, diffusivity-porosity, etc.) rather than the detailed transport information are more required. Therefore, the analysis related to Fig. 10 indicates that one can predict the intrinsic permeability and the Knudsen diffusivity of a porous medium, and then use Eq. (26) to determine the apparent permeability. Such a scheme avoid the tedious numerical schemes required for accurately predict the detailed fluid flow filed in microchannels or micro porous media [52, 53].

Fig. 11(a) shows the correction factor of the reconstructed organic matters predicted by our LB simulations. For calculating the correction factor, pressure and temperature are set as 2500 psi (17236893Pa) and 323K, under which the dynamic viscosity is $1.73\times10^{-5}$ Pa s. In Fig. 11(a), as the mean pore diameter decreases or the overlap tolerance reduces, the correction factor becomes larger, indicating that the Knudsen diffusion plays a more significant role compared to viscous flow. Most of values of the correct factor are in the range of 1.3~2 with intrinsic permeability in the range of $2\times10^{-4}$~$10^{-2}$ md. Particularly, for the case with $<d>$=30nm and tolerance =0.1, values of the correction factor for low porosity are greater than 10 and would be as large as 50, implying that the dominant mass transport mechanism is Knudsen diffusion.

Fig. 11(b) shows the comparisons between our simulation results and that predicted by various corrections. For the corrections listed in Table 1, porosity and intrinsic permeability obtained from our simulations are used to calculate $b_k$, and further to determine $f_c$ based on Eq. (18). For Klinkenberg's correction and Beskok and Karniadakis-Civan's correction, $Kn$ is required to calculate $f_c$. While mean free path can be directly determined by Eq. (29), the character pore size $r$ of the reconstructed organic matters is not easy to be obtained. In the present study, the following expression proposed by Herd et al. is used to calculate $r$ [44]

$$r = 8.886\times10^{-2}(k/\varepsilon)^{0.5} \tag{30}$$

The units of $r$ and $k$ are nm and mD, respectively. After $r$ is determined, $Kn$ can be calculated which can be further adopted in Klinkenberg's correction and Beskok and Karniadakis-Civan's correction to calculate $f_c$. As can be seen from Fig. 11(b), the corrections of Heid et al. and Jones and Owens greatly underestimate the correction factor. When calculating $b_k$ by taking the effects of porosity into consideration, the agreement between our simulation results and the corrections (Correction of Sampath and Keighin's and correction of Florence et al.) becomes better. The best agreement is observed for the Klinkenberg's correction and Beskok and Karniadakis-Civan's correction, both of which are based on the Knudsen number. Following Ziarani and Aguilera [51], the two corrections based on Knudsen number is called Knudsen's correction. By comparing different corrections with data from Mesaverde formation, Ziarani and Aguilere [51] concluded that the Knudsen's corrections perform the best. Our results are in line with that of Ziarani and Aguilere. While Ziarani and Aguilere found that Knudsen's corrections underestimate the apparent permeability for low permeability shales, it can be observed in Fig. 9(b) that Knudsen's corrections match with the simulation results well even under low permeability. Therefore, based on Ref. [51] as well as the present study, Knudsen's corrections combined with Eq. (30) are recommended to calculate the apparent permeability for organic matters.

## 5. Conclusion

In this work, Knudsen diffusion and fluid flow through the organic matters of shale matrix are simulated using the LBM. The nanoscale structures of the organic matters are reconstructed by randomly placing pore spheres into the domains. 150 samples of organic matters are reconstructed under different pore diameter, overlap tolerance and porosity. Effective Knudsen diffusivity and intrinsic permeability are predicted based on the concentration and velocity fields obtained from the LB simulation, and are compared with the values predicted the empirical equations. An equation is developed to determine the apparent permeability based on the effective Knudsen diffusivity and the intrinsic permeability. The calculated apparent permeability is also compared with various empirical corrections in the literature. The main conclusions of the present study are as follows.

1. The connectivity of the void space in the organic matters increases as the porosity increases. Under the same porosity, the connectivity of void space reduces as the pore diameter increases or the overlap tolerance increases.

2. The predicted Knudsen diffusivity ranges from $10^{-9}$ m$^2$ s$^{-1}$ to $10^{-6}$ m$^2$ s$^{-1}$. Bruggeman equation underestimates the tortuosity of the organic matters. Following the relationship $D_{eff} = D\varepsilon^{\alpha}$, based on the pore-scale simulation results in the present study, α in the range of 2.8~4.0 is recommended for calculating the effective diffusivity in the organic matters; and the lower the porosity and the overlap tolerance are, the higher the α is.

3. The predicted intrinsic permeability ranges $10^{-21}(10^{-6}mD) \sim 10^{-17}(10^{-2}mD)$ m$^2$. The Kozeny-Carman (KC) equation can be adopted to roughly estimate the intrinsic permeability of the organic matters for low or moderate porosity.

4. Transport behaviors of shale gas in organic matters are combined results of viscous fluid flow and Knudsen diffusion, which can be described by Eq. (25). Apparent permeability of the organic matters can be calculated based on the effective Knudsen diffusivity and intrinsic permeability. The corrections of only using intrinsic permeability to calculate the Klinkenberg's slippage factor (corrections of Heid et al. and Jones and Owens) greatly underestimates the apparent permeability. Corrections further taking into consideration the effects of porosity (Correction of Sampath and Keighin's and correction of Florence et al.) can improve the prediction. The Klinkenberg's correction and Beskok and Karniadakis-Civan's correction, both of which are based on the Knudsen number and thus are called Knudsen's corrections, match the best with our numerical simulations, which is in line with the experimental study in the literature [51]. Therefore, Knudsen's corrections are recommended to calculate the apparent permeability of the organic matters.

**Acknowledgement**

The authors acknowledge the support of LANL's LDRD Program, Institutional Computing Program, National Nature Science Foundation of China (No. 51406145 and 51136004) and NNSFC international-joint key project (No. 51320105004).

**Appendix A**

The equilibrium velocity moments $\mathbf{m}^{eq}$ are given as followes [29]

$$\mathbf{m}_0^{eq} = \rho \tag{A1}$$

$$\mathbf{m}_1^{eq} = -11\rho + 19\frac{\mathbf{j}\cdot\mathbf{j}}{\rho_0}, \quad \mathbf{m}_2^{eq} = 3\rho - \frac{11}{2}\frac{\mathbf{j}\cdot\mathbf{j}}{\rho_0} \tag{A2}$$

$$\mathbf{m}_3^{eq} = j_x, \quad \mathbf{m}_4^{eq} = -\frac{2}{3}j_x \tag{A3}$$

$$\mathbf{m}_5^{eq} = j_y, \quad \mathbf{m}_6^{eq} = -\frac{2}{3}j_y \tag{A4}$$

$$\mathbf{m}_7^{eq} = j_z, \quad \mathbf{m}_8^{eq} = -\frac{2}{3}j_z \tag{A5}$$

$$\mathbf{m}_9^{eq} = \frac{3j_x^2 - \mathbf{j}\cdot\mathbf{j}}{\rho_0}, \quad \mathbf{m}_{10}^{eq} = -\frac{3j_x^2 - \mathbf{j}\cdot\mathbf{j}}{2\rho_0} \tag{A6}$$

$$\mathbf{m}_{11}^{eq} = \frac{j_y^2 - j_z^2}{\rho_0}, \quad \mathbf{m}_{12}^{eq} = -\frac{j_y^2 - j_z^2}{2\rho_0} \tag{A7}$$

$$\mathbf{m}_{13}^{eq} = \frac{j_x j_y}{\rho_0}, \quad \mathbf{m}_{14}^{eq} = \frac{j_y j_z}{\rho_0}, \quad \mathbf{m}_{15}^{eq} = \frac{j_x j_z}{\rho_0} \tag{A8}$$

$$\mathbf{m}_{16-18}^{eq} = 0 \tag{A9}$$

$\rho_0$ is the mean density of the fluid, which is employed to reduce the compressibility effects of the model [29, 30].

The transformation matrix $\mathbf{Q}$ is as follows

|    | 0  | 1   | 2   | 3   | 4   | 5   | 6   | 7  | 8  | 9  | 10 | 11 | 12 | 13 | 14 | 15 | 16 | 17 | 18 |
|----|----|-----|-----|-----|-----|-----|-----|----|----|----|----|----|----|----|----|----|----|----|----|
| 0  | 1  | 1   | 1   | 1   | 1   | 1   | 1   | 1  | 1  | 1  | 1  | 1  | 1  | 1  | 1  | 1  | 1  | 1  | 1  |
| 1  | -30| -11 | -11 | -11 | -11 | -11 | -11 | 8  | 8  | 8  | 8  | 8  | 8  | 8  | 8  | 8  | 8  | 8  | 8  |
| 2  | 12 | -4  | -4  | -4  | -4  | -4  | -4  | 1  | 1  | 1  | 1  | 1  | 1  | 1  | 1  | 1  | 1  | 1  | 1  |
| 3  | 0  | 1   | -1  | 0   | 0   | 0   | 0   | 1  | -1 | 1  | -1 | 1  | -1 | 1  | -1 | 0  | 0  | 0  | 0  |
| 4  | 0  | -4  | 4   | 0   | 0   | 0   | 0   | 1  | -1 | 1  | -1 | 1  | -1 | 1  | -1 | 0  | 0  | 0  | 0  |
| 5  | 0  | 0   | 0   | 1   | -1  | 0   | 0   | 1  | 1  | -1 | -1 | 0  | 0  | 0  | 0  | 1  | -1 | 1  | -1 |
| 6  | 0  | 0   | 0   | -4  | 4   | 0   | 0   | 1  | 1  | -1 | -1 | 0  | 0  | 0  | 0  | 1  | -1 | 1  | -1 |
| 7  | 0  | 0   | 0   | 0   | 0   | 1   | -1  | 0  | 0  | 0  | 0  | 1  | 1  | -1 | -1 | 1  | 1  | -1 | -1 |
| 8  | 0  | 0   | 0   | 0   | 0   | -4  | 4   | 0  | 0  | 0  | 0  | 1  | 1  | -1 | -1 | 1  | 1  | -1 | -1 |
| 9  | 0  | 2   | 2   | -1  | -1  | -1  | -1  | 1  | 1  | 1  | 1  | 1  | 1  | 1  | 1  | -2 | -2 | -2 | -2 |
| 10 | 0  | -4  | -4  | 2   | 2   | 2   | 2   | 1  | 1  | 1  | 1  | 1  | 1  | 1  | 1  | -2 | -2 | -2 | -2 |
| 11 | 0  | 0   | 0   | 1   | 1   | -1  | -1  | 1  | 1  | 1  | 1  | -1 | -1 | -1 | -1 | 0  | 0  | 0  | 0  |
| 12 | 0  | 0   | 0   | -2  | -2  | 2   | 2   | 1  | 1  | 1  | 1  | -1 | -1 | -1 | -1 | 0  | 0  | 0  | 0  |
| 13 | 0  | 0   | 0   | 0   | 0   | 0   | 0   | 1  | -1 | -1 | 1  | 0  | 0  | 0  | 0  | 0  | 0  | 0  | 0  |
| 14 | 0  | 0   | 0   | 0   | 0   | 0   | 0   | 0  | 0  | 0  | 0  | 0  | 0  | 0  | 0  | 1  | -1 | -1 | 1  |
| 15 | 0  | 0   | 0   | 0   | 0   | 0   | 0   | 0  | 0  | 0  | 0  | 1  | -1 | -1 | 1  | 0  | 0  | 0  | 0  |
| 16 | 0  | 0   | 0   | 0   | 0   | 0   | 0   | 1  | -1 | 1  | -1 | -1 | 1  | -1 | 1  | 0  | 0  | 0  | 0  |
| 17 | 0  | 0   | 0   | 0   | 0   | 0   | 0   | -1 | -1 | 1  | 1  | 0  | 0  | 0  | 0  | 1  | -1 | 1  | -1 |
| 18 | 0 | 0 | 0 | 0 | 0 | 0 | 0 | 0 | 0 | 0 | 0 | 1 | 1 | -1 | -1 | -1 | -1 | 1 | 1 |

And its inverse matrix $\mathbf{Q}^{-1}$ is as follows

| | 0 | 1 | 2 | 3 | 4 | 5 | 6 | 7 | 8 | 9 | 10 | 11 | 12 | 13 | 14 | 15 | 16 | 17 | 18 |
|---|---|---|---|---|---|---|---|---|---|---|---|---|---|---|---|---|---|---|---|
| 0 | 1/19 | -15/1197 | 1/21 | 0 | 0 | 0 | 0 | 0 | 0 | 0 | 0 | 0 | 0 | 0 | 0 | 0 | 0 | 0 | 0 |
| 1 | 1/19 | -11/2394 | -1/63 | 0.1 | -0.1 | 0 | 0 | 0 | 0 | 1/18 | -1/18 | 0 | 0 | 0 | 0 | 0 | 0 | 0 | 0 |
| 2 | 1/19 | -11/2394 | -1/63 | -0.1 | 0.1 | 0 | 0 | 0 | 0 | 1/18 | -1/18 | 0 | 0 | 0 | 0 | 0 | 0 | 0 | 0 |
| 3 | 1/19 | -11/2394 | -1/63 | 0 | 0 | 0.1 | -0.1 | 0 | 0 | -1/36 | 1/36 | 1/12 | -1/12 | 0 | 0 | 0 | 0 | 0 | 0 |
| 4 | 1/19 | -11/2394 | -1/63 | 0 | 0 | -0.1 | 0.1 | 0 | 0 | -1/36 | 1/36 | 1/12 | -1/12 | 0 | 0 | 0 | 0 | 0 | 0 |
| 5 | 1/19 | -11/2394 | -1/63 | 0 | 0 | 0 | 0 | 0.1 | -0.1 | -1/36 | 1/36 | -1/12 | 1/12 | 0 | 0 | 0 | 0 | 0 | 0 |
| 6 | 1/19 | -11/2394 | -1/63 | 0 | 0 | 0 | 0 | -0.1 | 0.1 | -1/36 | 1/36 | -1/36 | 1/36 | 0 | 0 | 0 | 0 | 0 | 0 |
| 7 | 1/19 | 4/1197 | 1/252 | 0.1 | 1/40 | 0.1 | 1/40 | 0 | 0 | 1/36 | 1/72 | 1/12 | 1/24 | 1/4 | 0 | 0 | 1/8 | -1/8 | 0 |
| 8 | 1/19 | 4/1197 | 1/252 | -0.1 | -1/40 | 0.1 | 1/40 | 0 | 0 | 1/36 | 1/72 | 1/12 | 1/24 | -1/4 | 0 | 0 | -1/8 | -1/8 | 0 |
| 9 | 1/19 | 4/1197 | 1/252 | 0.1 | 1/40 | -0.1 | -1/40 | 0 | 0 | 1/36 | 1/72 | 1/12 | 1/24 | -1/4 | 0 | 0 | 1/8 | 1/8 | 0 |
| 10 | 1/19 | 4/1197 | 1/252 | -0.1 | -1/40 | -0.1 | -1/40 | 0 | 0 | 1/36 | 1/72 | 1/12 | 1/24 | 1/4 | 0 | 0 | -1/8 | 1/8 | 0 |
| 11 | 1/19 | 4/1197 | 1/252 | 0.1 | 1/40 | 0 | 0 | 0.1 | 1/40 | 1/36 | 1/72 | -1/12 | -1/24 | 0 | 0 | 1/4 | -1/8 | 0 | 1/8 |
| 12 | 1/19 | 4/1197 | 1/252 | -0.1 | -1/40 | 0 | 0 | 0.1 | 1/40 | 1/36 | 1/72 | -1/12 | -1/24 | 0 | 0 | -1/4 | 1/8 | 0 | 1/8 |
| 13 | 1/19 | 4/1197 | 1/252 | 0.1 | 1/40 | 0 | 0 | -0.1 | -1/40 | 1/36 | 1/72 | -1/12 | -1/24 | 0 | 0 | -1/4 | -1/8 | 0 | -1/8 |
| 14 | 1/19 | 4/1197 | 1/252 | -0.1 | -1/40 | 0 | 0 | -0.1 | -1/40 | 1/36 | 1/72 | -1/12 | -1/24 | 0 | 0 | 1/4 | 1/8 | 0 | -1/8 |
| 15 | 1/19 | 4/1197 | 1/252 | 0 | 0 | 0.1 | 1/40 | 0.1 | 1/40 | -1/18 | -1/36 | 0 | 0 | 0 | 1/4 | 0 | 0 | 1/8 | -1/8 |
| 16 | 1/19 | 4/1197 | 1/252 | 0 | 0 | -0.1 | -1/40 | 0.1 | 1/40 | -1/18 | -1/36 | 0 | 0 | 0 | -1/4 | 0 | 0 | -1/8 | -1/8 |
| 17 | 1/19 | 4/1197 | 1/252 | 0 | 0 | 0.1 | 1/40 | -0.1 | -1/40 | -1/18 | -1/36 | 0 | 0 | 0 | -1/4 | 0 | 0 | 1/8 | 1/8 |
| 18 | 1/19 | 4/1197 | 1/252 | 0 | 0 | -0.1 | -1/40 | -0.1 | -1/40 | -1/18 | -1/36 | 0 | 0 | 0 | 1/4 | 0 | 0 | -1/8 | 1/8 |

**Author contributions**

L. C. wrote the main manuscript and prepared figures 1-8. L.Z. helped to reconstruct the structures of shales. Q.J.K. and W.Q.T. supervised the theoretical analysis and writing. All authors reviewed the manuscript.

**Additional information**

Competing financial interests: The authors declare no competing financial interests.

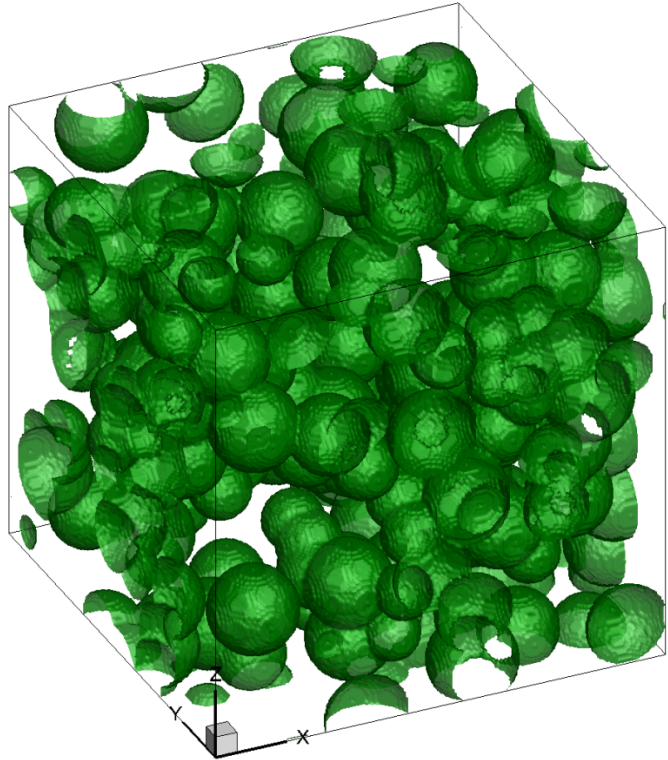

Fig. 1 A 3D representation of a reconstructed geometry with mean diameter of 45nm, porosity of 0.4 and overlap tolerance of 0.2. The green spheres are the pore spheres. The computational domain is a cube with size of 300×300×300nm, and is discretized by 200×200×200 lattices, leading to a resolution of 1.5nm per lattice.

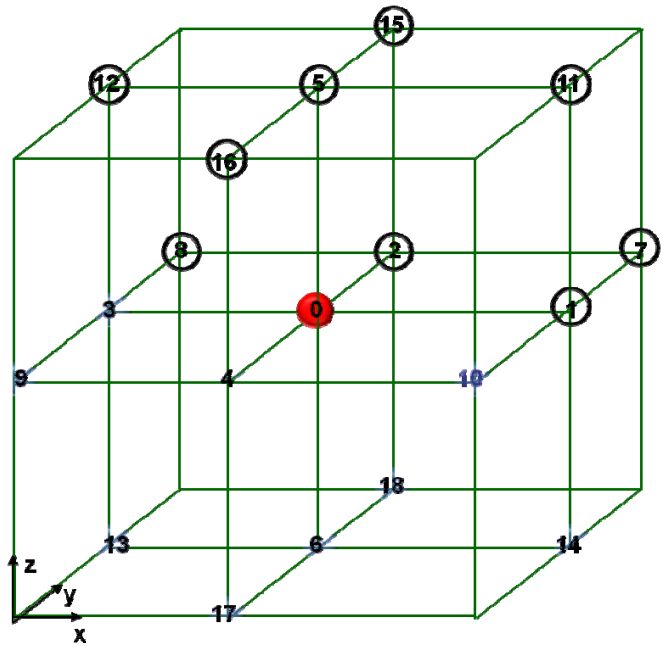

Fig.2 D3Q19 lattice model

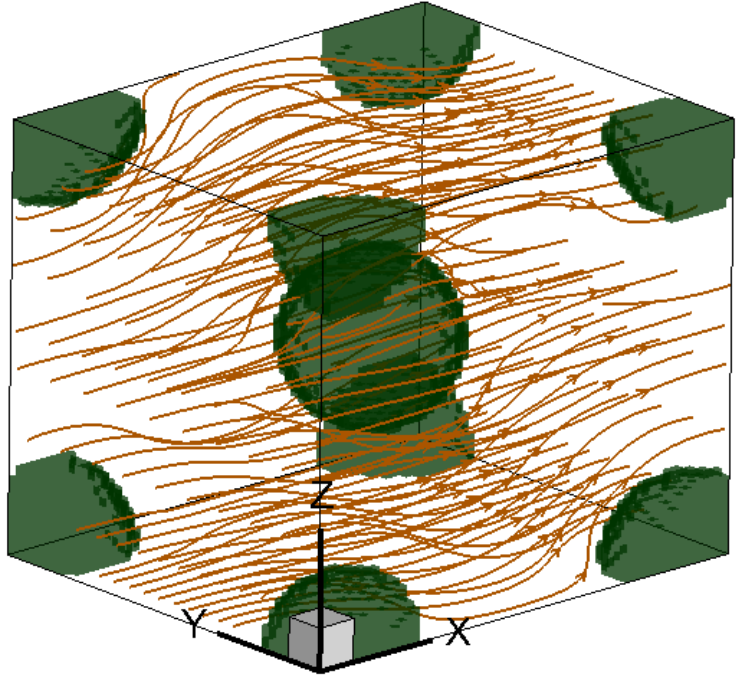

(a) The simulated streamlines inside the arrays with a porosity of 0.915.

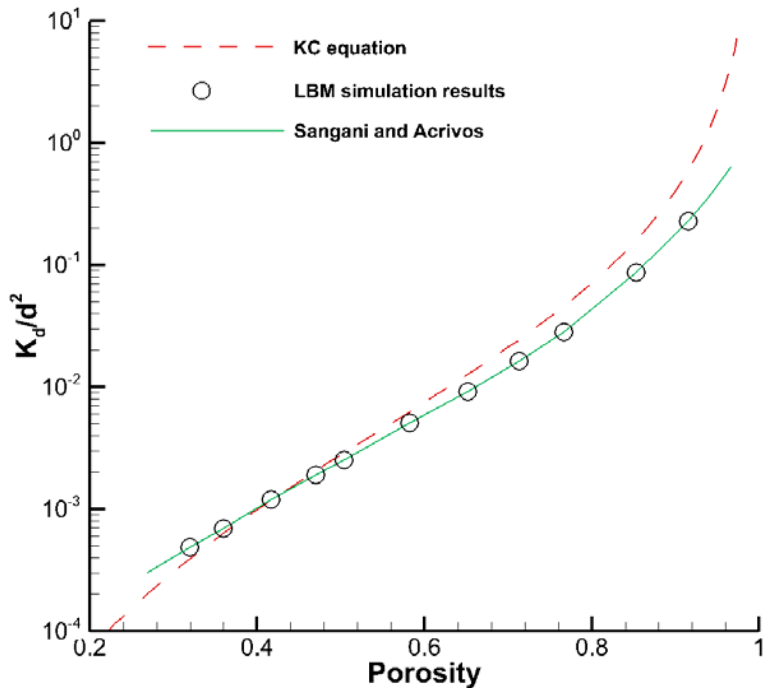

(b) Comparison of the LBM simulated intrinsic permeability with the KC relation and the analytical solution

Fig.3 Fluid flow and permeability for periodic BBC arrays

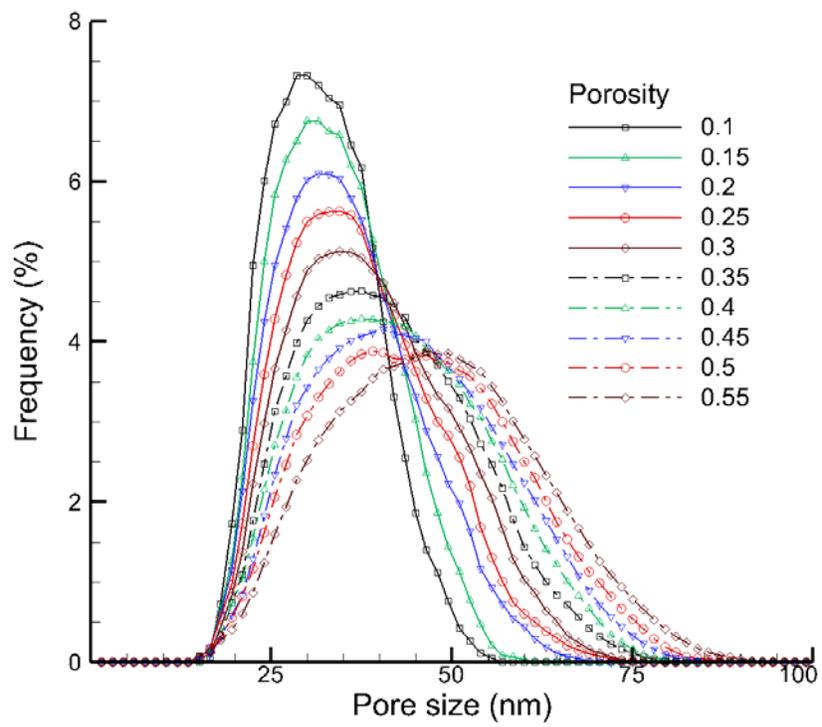

Fig. 4 The PSD for the reconstructed organic matters with $<d>$=45nm and $\xi$=0.1 for different porosities

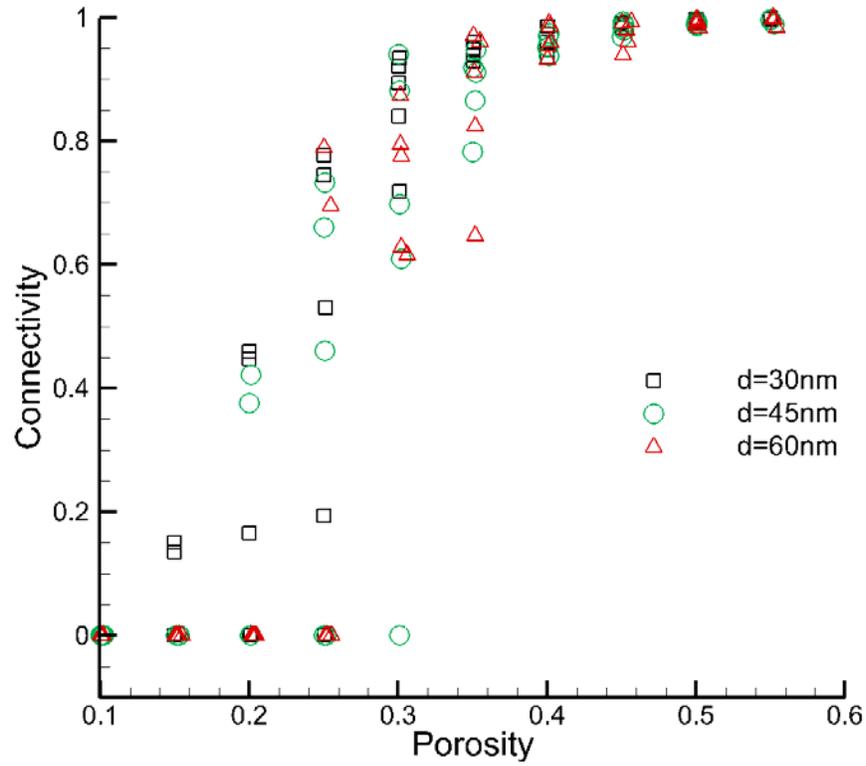

(a) Effects of pore sphere diameter

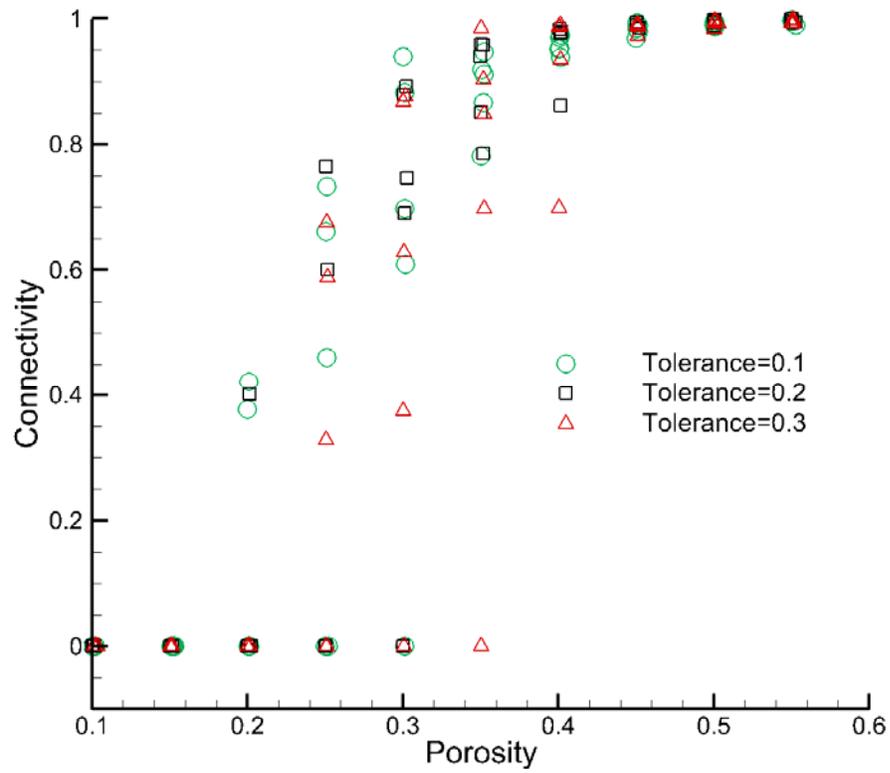

(b) Effects of overlap tolerance

Fig. 5 The connectivity of void space for different reconstructed organic matters

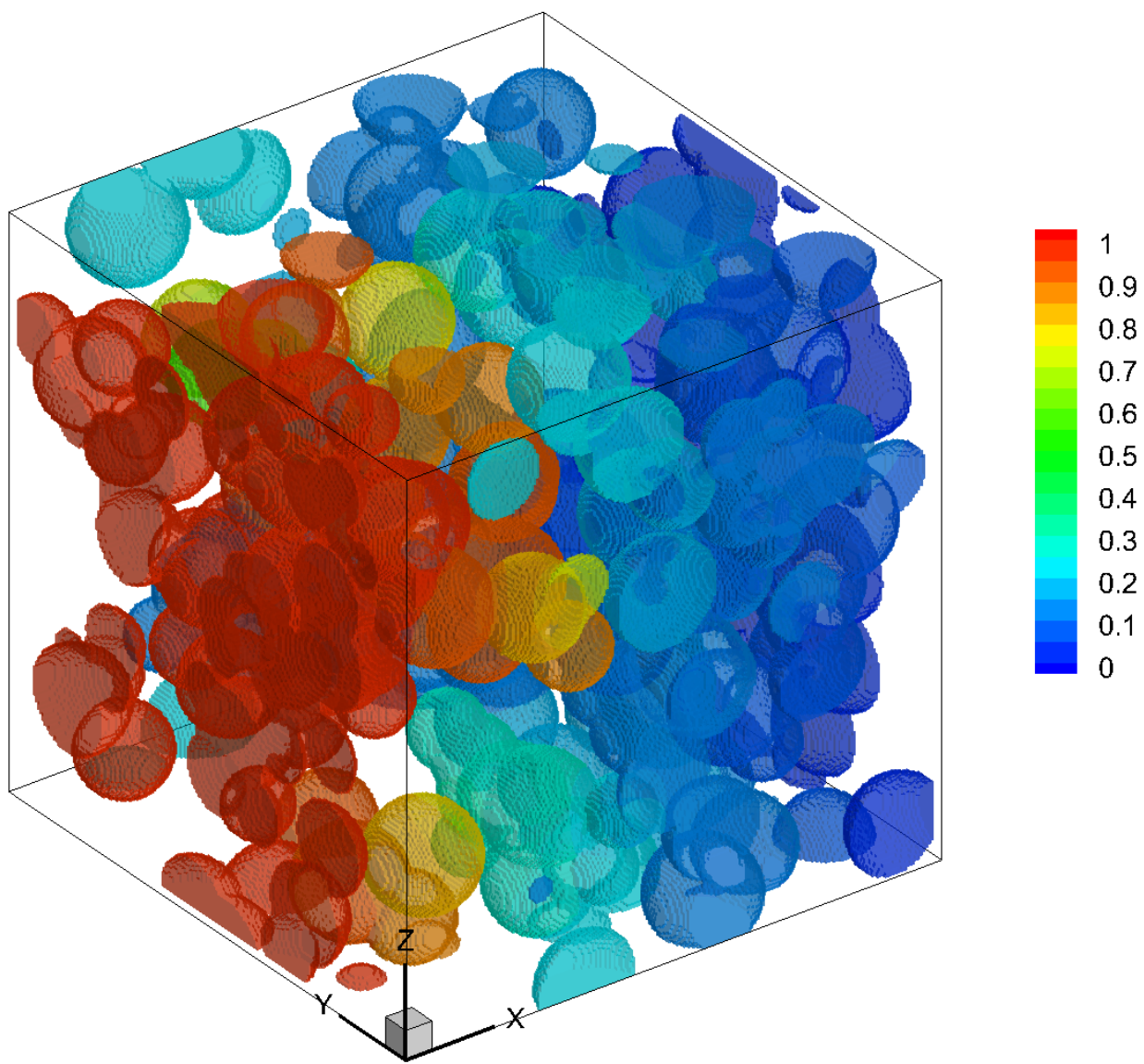

Fig. 6 Concentration distribution within the reconstructed organic matter with mean diameter of 45nm, porosity of 0.4 and overlap tolerance of 0.2.

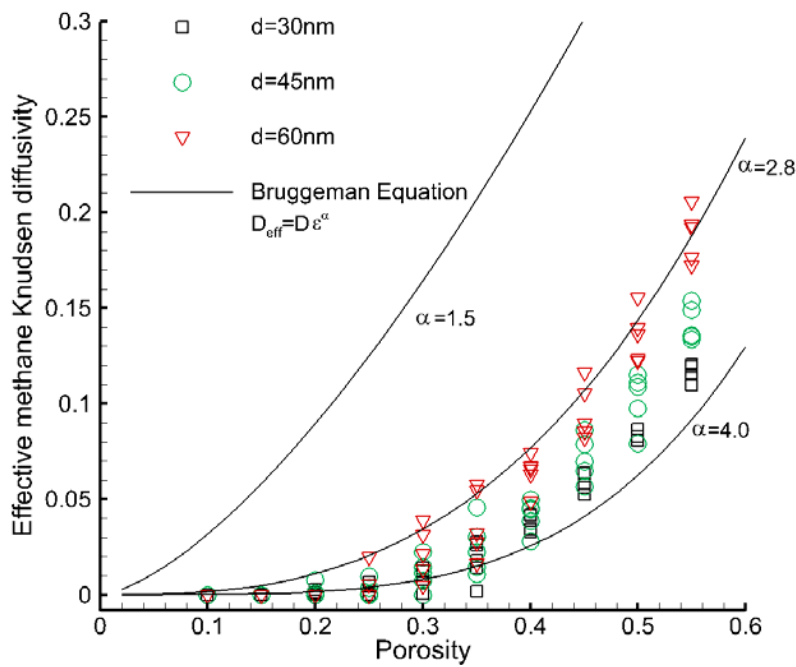

(a) Effects of pore sphere size

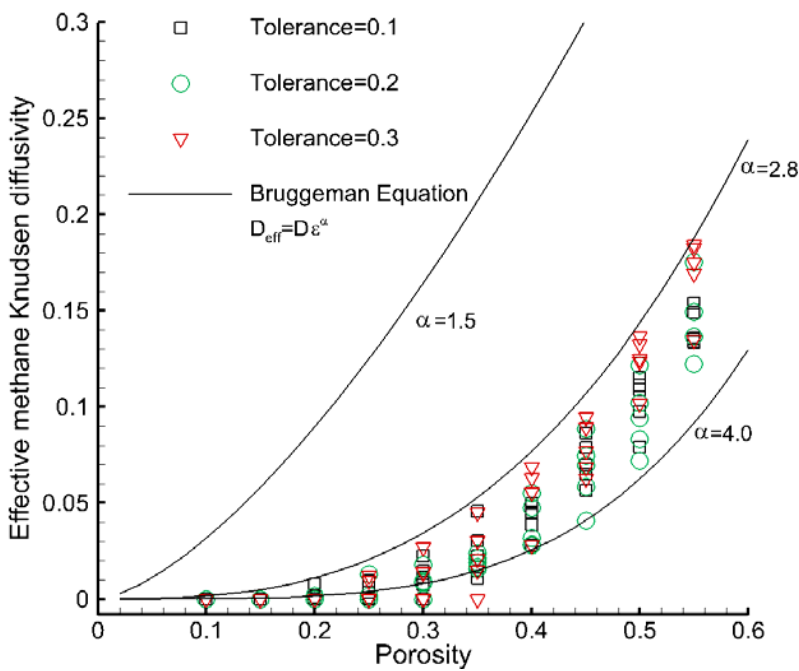

(b) Effects of overlap tolerance

Fig. 7 Effective Knudsen diffusivity for the reconstructed organic matters. The diffusivity is normalized by the Knudsen diffusivity with $d$=60nm and $T$=323. The predicted Knudsen diffusivity ranges from $10^{-8}$ $m^2$ $s^{-1}$ to $10^{-6}$ $m^2$ $s^{-1}$. Revised Bruggeman equation with α ranges 2.8~4.0 are in good agreement with values predicted by our simulations.

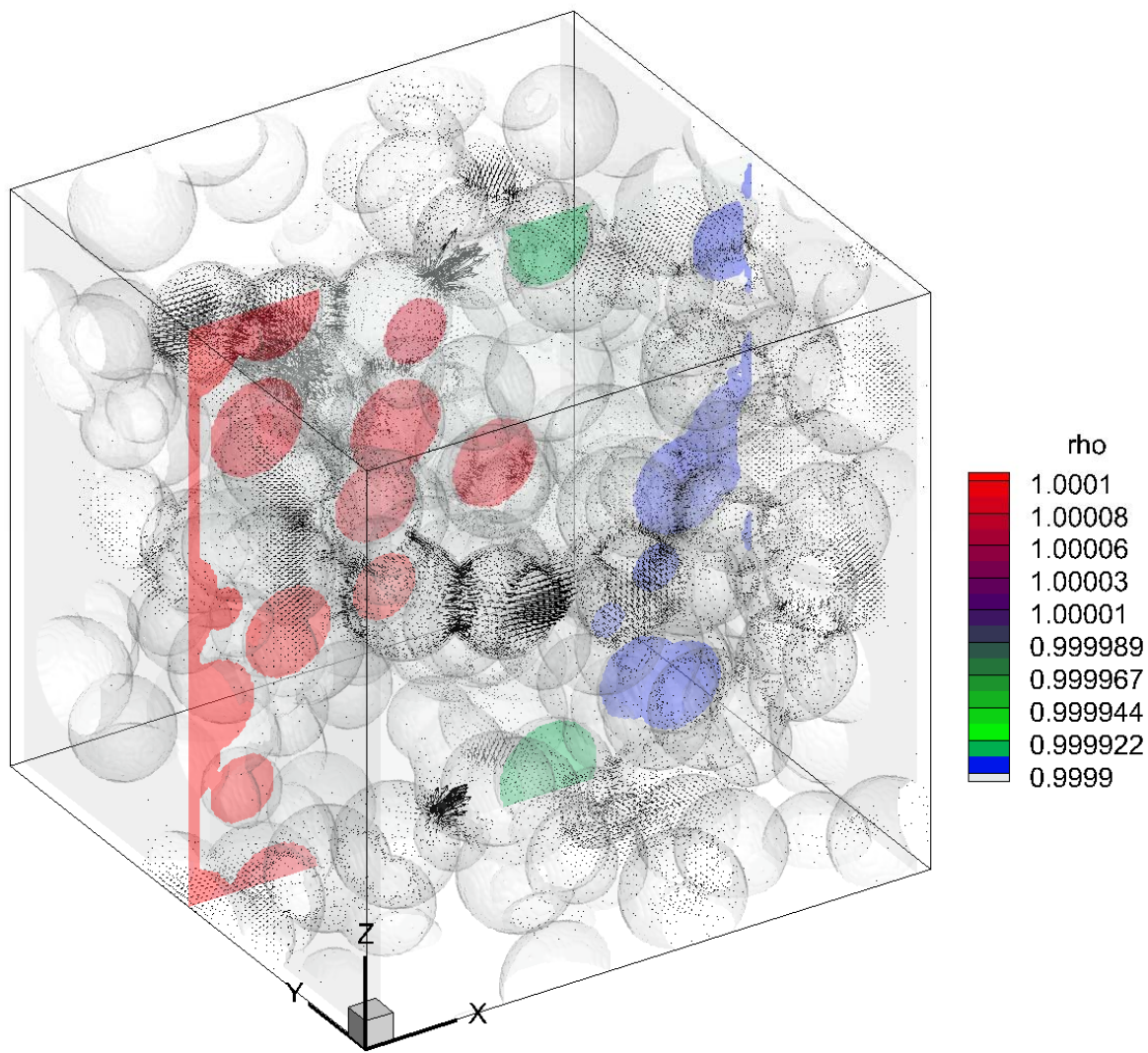

Fig. 8 Pressure and velocity vector distributions within the reconstructed organic matters with mean diameter of 45nm, porosity of 0.4 and overlap tolerance of 0.2.

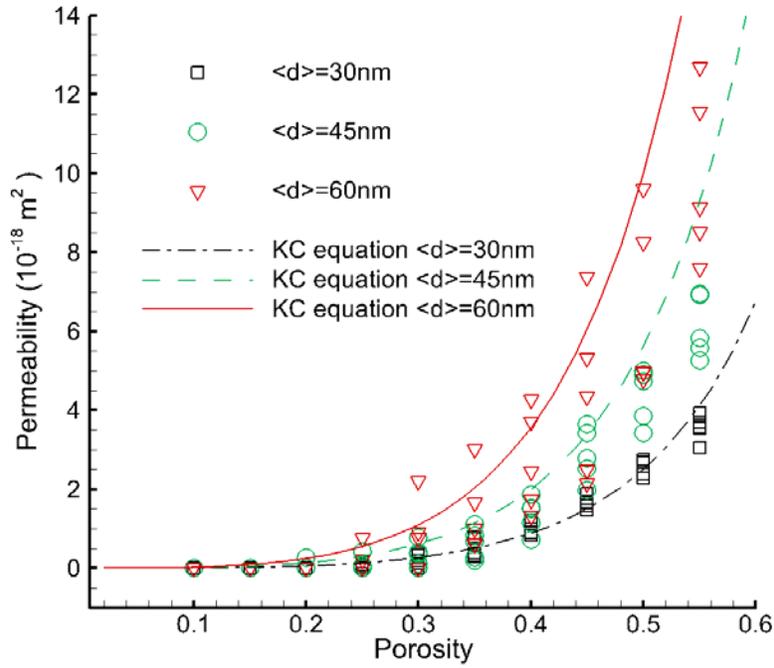

(a) Effects of pore sphere size

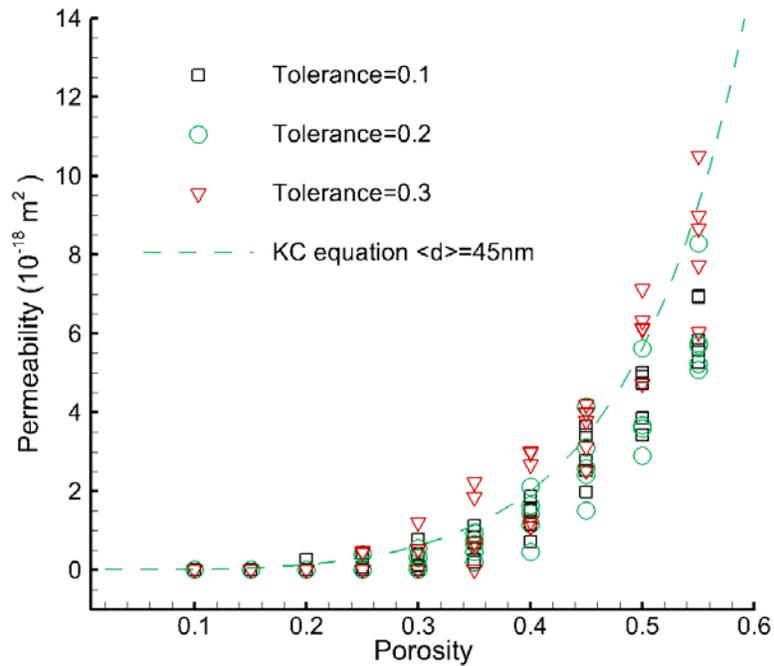

(b) Effects of overlap tolerance

Fig. 9 Intrinsic permeability for the reconstructed organic matters. The intrinsic permeability predicted ranges $10^{-21}(10^{-6}\text{mD}) \sim 10^{-17}(10^{-2}\text{mD})$ m$^2$. The value of permeability predicted by KC equation presents acceptable agreement with that obtained from our nanoscale simulations.

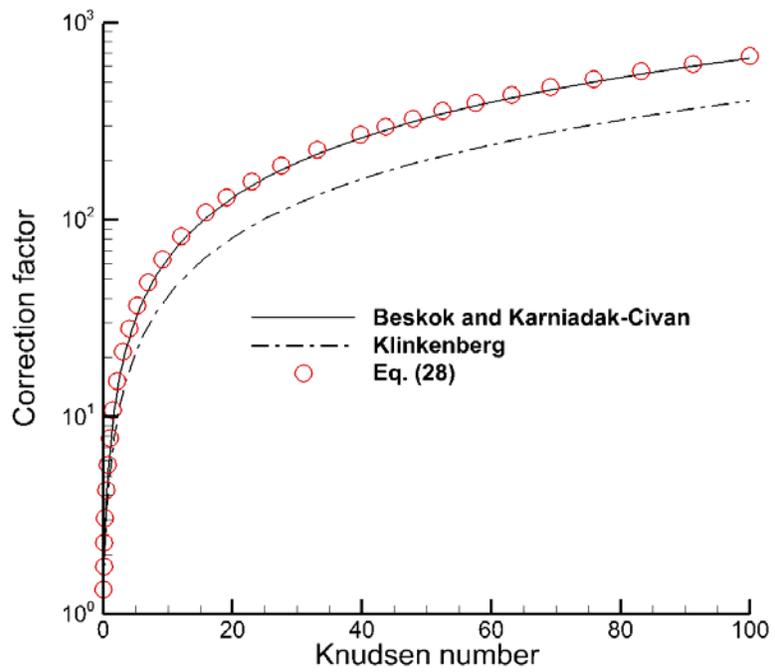

Fig. 10 Correction factor between intrinsic and apparent permeability predicted by different corrections.

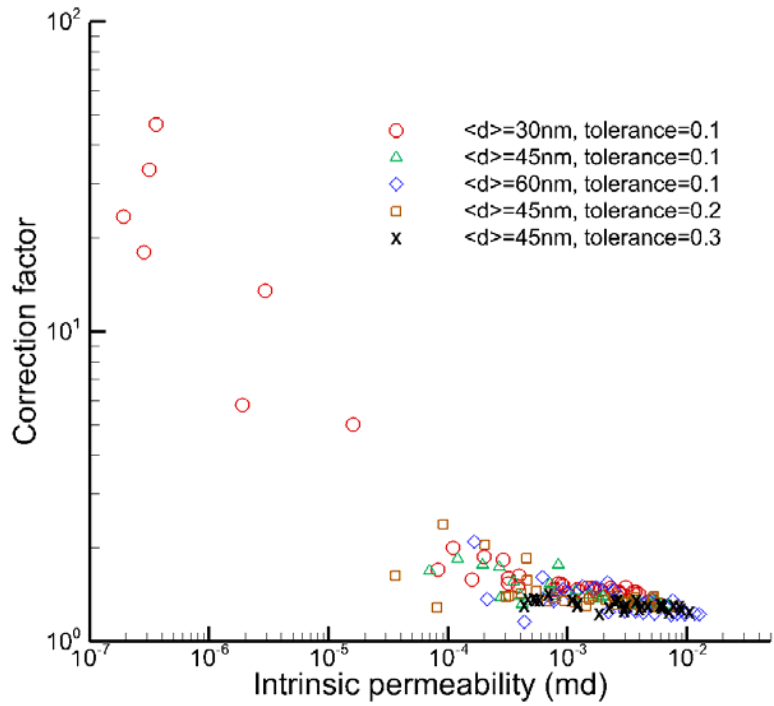

(a) The correction factor based on the numerical results

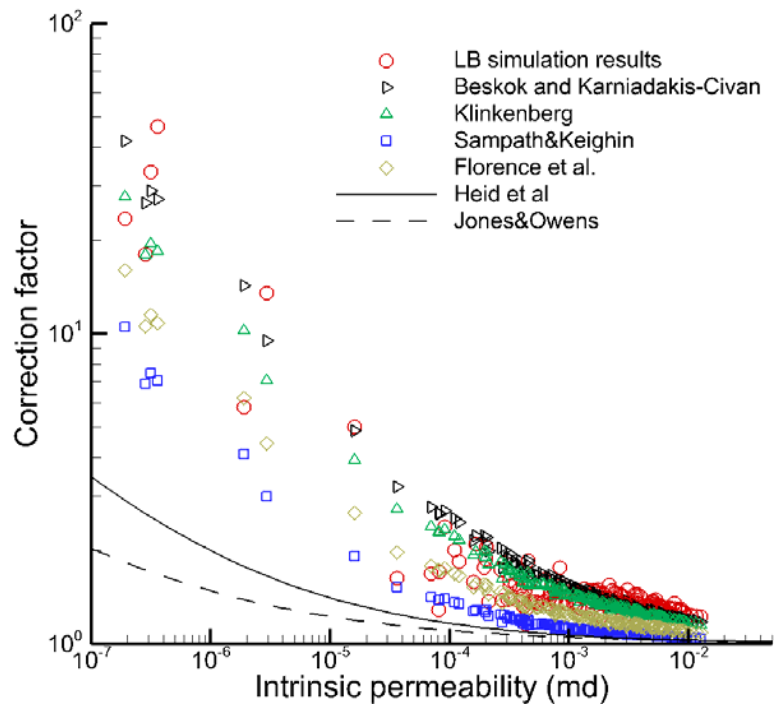

(b) Comparison between the numerical results and different corrections in the literature

Fig. 11 The correction factor between apparent permeability and intrinsic permeability